\documentclass[12pt,preprint]{aastex}

\slugcomment{ApJL, in press}

\shorttitle{Accretion during black hole mergers}
\shortauthors{Armitage \& Natarajan}

\begin{document}

\title{Accretion during the merger of supermassive black holes}

\author{Philip J. Armitage\altaffilmark{1}}
\affil{School of Physics and Astronomy, University of St Andrews, Fife, 
       KY16 9SS, UK}
\altaffiltext{1}{Present address: JILA, University of Colorado, 440 UCB, 
        Boulder, CO 80309-0440}       
\email{pja3@st-andrews.ac.uk}

\and

\author{Priyamvada Natarajan}
\affil{Astronomy Department, Yale University, P.O. Box 208101, 
       New Haven, CT 06520-8101, USA}
\email{priya@astro.yale.edu}

\begin{abstract}
We study the evolution of disk accretion during the merger of supermassive 
black hole binaries in galactic nuclei. In hierarchical galaxy formation 
models, the most common binaries are likely to arise from minor galactic 
mergers, and have unequal mass black holes. Once such a binary becomes 
embedded in an accretion disk at a separation $a \sim 0.1 \ {\rm pc}$, 
the merger proceeds in two distinct phases. During the first phase, the loss 
of orbital angular momentum to the gaseous disk shrinks the binary on a 
timescale of $\sim 10^7 \ {\rm yr}$. The accretion 
rate onto the primary black hole is not increased, and can be 
substantially reduced, during this disk-driven migration. At 
smaller separations, gravitational radiation becomes the 
dominant angular momentum loss process, and any gas trapped inside 
the orbit of the secondary is driven inwards by the 
inspiralling black hole. The implied 
accretion rate just prior to coalescence exceeds the Eddington limit, so 
the final merger is likely to occur within a common envelope 
formed from the disrupted inner disk, and be accompanied by high velocity ($\sim 10^4 \ 
{\rm kms}^{-1}$) outflows.
\end{abstract}


\section{Introduction}
A large fraction of galaxies in the local Universe harbor supermassive 
black holes \citep{magorrian98}, with the black hole mass  
correlating well with the velocity dispersion of the galaxy \citep{ferrarese2000,gebhardt2000}.
One explanation for this correlation is that black holes, like galaxies 
themselves, have grown in mass through the hierarchical merging of smaller 
progenitors \citep{haehnelt2000,cv2000,menou2001}. This implies  
that the frequent galactic mergers at high redshift could also involve 
the inspiral, and probably eventual coalescence, of supermassive black holes. 
Typically, these binaries will pair black holes of disparate masses, since 
minor galactic mergers are more common than major mergers in 
hierarchical models of galaxy formation \citep{haehnelt2000}.

For a black hole binary with masses of $10^8 \ M_\odot$ and $10^6 \ M_\odot$, 
gravitational radiation will lead to merger within a Hubble time if the 
separation is $\lesssim 10^{-2} \ {\rm pc}$. Reaching such a small 
separation, following a galactic merger, requires that the black 
holes lose almost all of their orbital angular momentum to stars, dark matter,  
or gas. Angular momentum loss to stars occurs readily at 
large radii, but slows down at pc-scale 
separations if the supply of stars on low angular momentum 
orbits, which can interact with the binary, becomes depleted 
\citep{begelman80,mikkola92,makino97,quinlan97}. The extent 
of this loss cone depletion remains a subject of active 
debate \citep{milo2001,yu2001,zhao01}, with recent work 
suggesting that stellar dynamical processes may 
suffice to bring binaries into the gravitational radiation dominated regime.
Irrespective of these uncertainties, gas, if it is present at 
small radii in the nucleus, is likely to be the most efficient 
sink of binary angular momentum \citep{begelman80,pringle91,ivanov99,gould2000}.
In this {\em letter}, we study the role of gas in hastening 
the merger process, and the influence of the binary on the accretion 
rate and luminosity of the black holes. 

\section{Timescales for black hole mergers}
When the mass ratio of the binary, $q = M_2 / M_1 \ll 1$, the 
interaction of the lower mass hole with the disk is identical 
to that of a Jupiter mass planet in a protoplanetary disk (ignoring 
relativistic effects due to black hole spin). This regime, which 
\citet{gould2000} described as `planet-like', has been 
extensively studied \citep{goldreich80,lin86}.  
The secondary black hole will open a gap in the 
disk around the primary, and exchange angular 
momentum with the gas via gravitational torques. If the disk mass is 
at least of the same order of magnitude as the secondary, 
orbital migration then occurs on the viscous timescale 
of the accretion disk \citep{syer95,ivanov99}. The radial 
velocity of fluid in the disk is given by $v_r = -3 \nu / 2 r$, 
where $\nu$ is the `viscosity' used to approximate angular 
momentum transport within the disk at radius $r$. Using the 
\citet{shakura73} $\alpha$ prescription, $\nu = \alpha c_s h$, 
where $c_s$ is the disk sound speed and $h$ the vertical scale 
height, the rate 
of migration is approximately 
\begin{equation} 
 \dot{a}_{\rm visc} \simeq - {3 \over 2} \left( {h \over r} \right)^2 
 \alpha v_K
\label{eq_visc2}
\end{equation}
where $v_K$ is the Keplerian velocity. Equating this to the 
rate of inspiral due to gravitational radiation, which for 
circular orbits is
\begin{equation}
 \dot{a}_{\rm GW} = - { {64 G^3 M_1 M_2 (M_1 + M_2)} \over 
 {5 c^5 a^3} }, 
\label{eq_gw}
\end{equation}  
we obtain the critical semi-major axis within which gravitational 
radiation is the dominant process leading to orbital decay. For 
$q \ll 1$ we find
\begin{equation}
 a_{\rm crit} = \left( {128 \over 5} \right)^{2/5}  
 \left( {h \over r} \right)^{-4/5} \alpha^{-2/5} 
 q^{2/5} \left( { {G M_1} \over c^2} \right).
\label{eq_acrit}
\end{equation} 
This transition radius between disk-driven migration and 
gravitational wave inspiral scales with the gravitational 
radius of the primary black hole, and decreases for more 
extreme mass ratios. For fiducial parameters of $(h/r) \approx 10^{-2}$, 
$\alpha = 10^{-2}$, and $q= 10^{-2}$, we obtain $a_{\rm crit} \sim 10^2 \  
GM_1 / c^2$, or around $3.5 \times 10^{-3} \ {\rm pc}$ for a primary 
black hole mass of $5 \times 10^8 \ M_\odot$.

Once gravitational radiation becomes significant, the $\dot{a}_{\rm GW} 
\propto a^{-3}$ dependence of equation (\ref{eq_gw}) ensures that 
disk torques rapidly become negligible. The merger timescale from 
$a_{\rm crit}$ is 
\begin{equation}
 t_{\rm merge} = { 1 \over {2 c} } \left( {128 \over 5} \right)^{3/5} 
 \left( {h \over r} \right)^{-16/5} \alpha^{-8/5} q^{3/5} 
 \left( { {G M_1} \over c^2} \right),
\label{eq_tmerge}
\end{equation} 
which is around $10^5 \ {\rm yr}$ for the parameters above. 
During this final stage of orbital decay, gravitational 
torques will still extract angular momentum from the 
disk {\em interior} to the orbit, maintaining a gap 
and forcing the inner disk to accrete more rapidly. Simulations 
show that gap opening occurs on an orbital timescale, so there 
is no {\em dynamical} reason why this driven inflow should 
not continue until the inspiral velocity reaches a significant 
fraction of the Keplerian velocity. Adopting the criteria that 
$\dot{a}_{\rm GW} \lesssim 10^{-2} v_K$, a gap could remain 
until $a \simeq 17 q^{2/5} ( {GM_1 / c^2} ) \ll a_{\rm crit}$.
In practice, the thin disk approximation will 
instead break down when the accretion rate through the inner 
disk exceeds the Eddington limit. 

For mass ratios $q \ge 10^{-2}$, 
$a_{\rm crit}$ is substantially outside the marginally stable orbit.
However, we note that for smaller $q$, 
disk-driven migration may 
{\em always} be more important than gravitational radiation. This means that, if 
a geometrically thin accretion disk is present, low frequency  
gravitational waves would not be detectable from extreme mass ratio binaries, 
except during the final dynamical stage of the merger. Geometrically thick accretion 
flows, however, are expected to be much less efficient at driving 
migration \citep{narayan00}.

\section{Numerical merger model}

We model the migration of the secondary black hole in the 
planet-like regime \citep{gould2000} using an approximate, one dimensional treatment. 
We assume that the orbit remains circular, and solve for the coupled evolution 
of the disk surface density $\Sigma (r,t)$ and secondary semi-major axis $a$. 
The governing equation for the surface density evolution is \citep{lin86,trilling98}
\begin{equation}
 { {\partial \Sigma} \over {\partial t} } =
 { 1 \over r } { \partial \over {\partial r} }
 \left[ 3 r^{1/2} { \partial \over {\partial r} }
 \left( \nu \Sigma r^{1/2} \right) -
 { { 2 \Lambda \Sigma r^{3/2} } \over
 { (G M_1)^{1/2} } } \right].
\label{eqsigma}
\end{equation}   
The first term on the right-hand side describes the usual diffusive 
evolution of an accretion disk. The second 
term describes how the disk responds to the torque from the 
secondary, which is approximated as a fixed function of radius 
$\Lambda (r,a)$. We adopt
\begin{eqnarray}
 \Lambda & = & - { {f q^2 G M_1} \over {2 r} }
 \left( {r \over \Delta_p} \right)^4 \, \, \, \, r < a \nonumber \\
 \Lambda & = & { {f q^2 G M_1} \over {2 r} }
 \left( {a \over \Delta_p} \right)^4 \, \, \, \, r > a
\label{eqtorque}
\end{eqnarray}
where $f$ is a dimensionless normalization factor and $\Delta_p$ is 
given by $\Delta_p = {\rm max} ( h, \vert r - a \vert )$.
The rate of migration of the secondary is
\begin{equation}
 { { {\rm d} a } \over { {\rm d} t } } =
 - \left( { a \over {GM_1} } \right)^{1/2}
 \left( { {4 \pi} \over M_2 } \right)
 \int_{r_{\rm in}}^{r_{\rm out}}
 r \Lambda \Sigma {\rm d} r + \dot{a}_{\rm GW},
\end{equation}  
where the integral is taken over the entire disk.

\begin{figure}
\plotone{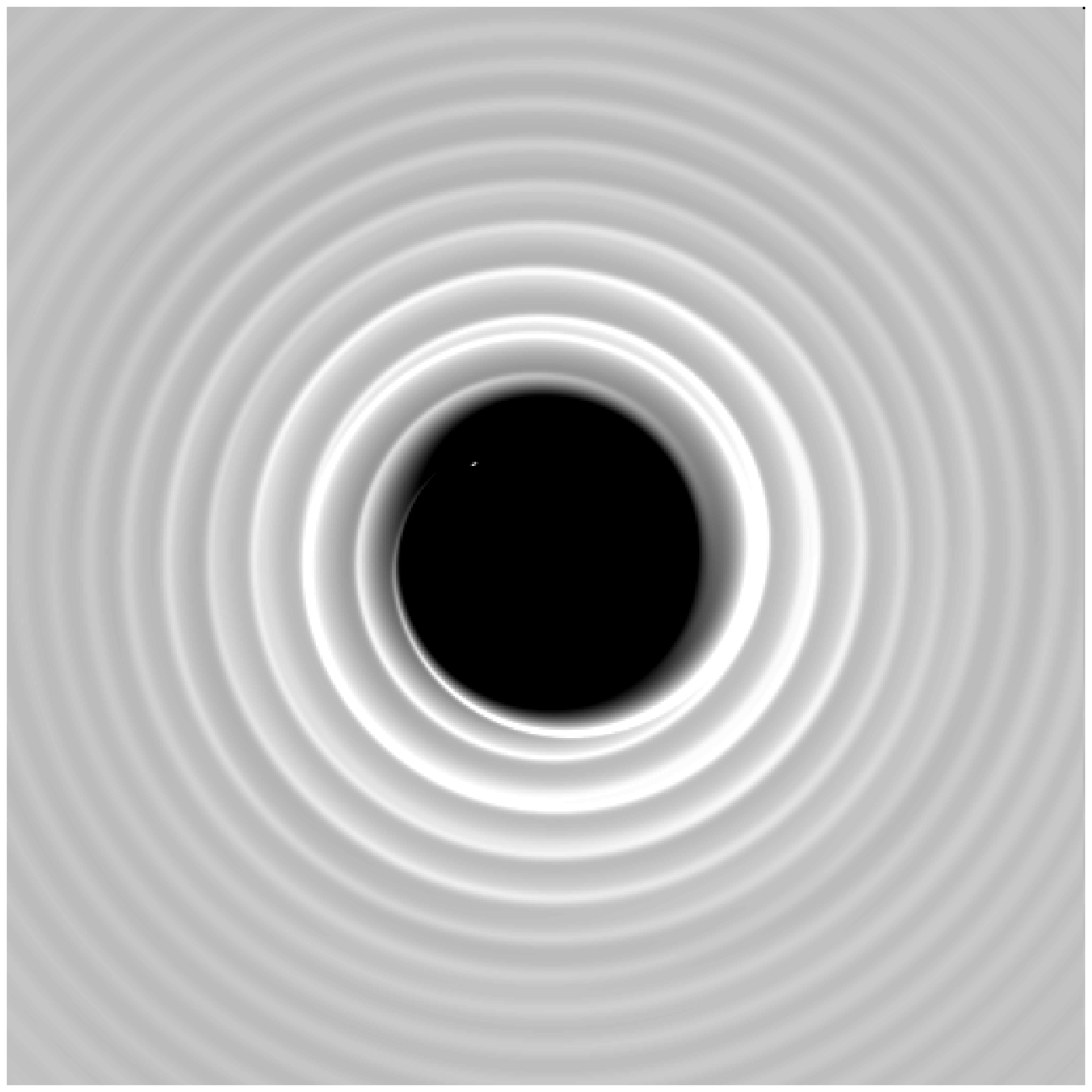}
\caption{The interaction between a low mass secondary black hole and 
	the outer accretion disk, simulated at high resolution in two 
	dimensions. The tightly wound spiral waves that mediate angular 
	momentum exchange between the secondary and the disk are 
	clearly seen.
	This run was for a mass ratio $q=0.01$, disk thickness 
	$(h/r) \simeq 0.07$, and a constant kinematic viscosity $\nu$ 
	appropriate to a disk with $\alpha \approx 10^{-2}$ at the radius 
	of the satellite. In 
	general, similar waves would also propagate through an 
	{\em inner} accretion disk, which was not included in this calculation.} 
\label{fig1}
\end{figure}

To fix $f$, and to help visualize the interaction, we have run 
high resolution ($400^2$ mesh points) simulations of the secondary-disk 
interaction using the ZEUS hydrodynamics code \citep{stone92}. Figure 
\ref{fig1} shows the results from a locally isothermal ($c_s = c_s(r)$ 
only) calculation, which was evolved for 400 
orbits of the secondary until a near steady-state was obtained. For 
both this run, with $(h/r) \simeq 0.07$, and for simulations of cooler disks 
with $(h/r) = 0.035, 0.02$, the edge of the gap 
lies close to the 2:1 orbital 
resonance. A low level of accretion onto the secondary from the outer 
disk is observed. In our 1D disk code described below, gaps of approximately 
the correct width are obtained 
using $f=10^{-2}$.

We complete the model by adopting a viscosity 
$\nu = 1.5 \times 10^{20} \ ( { r / {0.1 \ {\rm pc}} } )^{3/2} \ 
{\rm cm^2 s^{-1}}$,
where the prefactor is chosen to match the viscosity of a disk 
with $\alpha = 10^{-2}$ and $(h/r) = 0.01$ at 0.1~pc. The scaling 
with radius is model dependent and poorly known. We adopt a 
power-law with an index of 3/2 primarily for numerical convenience.

\subsection{Results for prompt mergers}

Figure 2 shows the evolution of a black hole binary with masses 
$M_1 = 5 \times 10^8 \ M_\odot$ and $M_2 = 10^7 \ M_\odot$, calculated 
by numerically integrating equation (\ref{eqsigma}) using an 
explicit method on a non-uniform grid. 800 mesh points were used 
to resolve the disk from $r_{\rm ms} = 1.4 \times 10^{-4} \ {\rm pc}$ 
out to $3 \ {\rm pc}$. 
The disk has a zero-torque boundary condition at the inner edge, 
and an initial surface density profile corresponding to a constant 
accretion rate of $1 \ M_\odot {\rm yr}^{-1}$ out to $1 \ {\rm pc}$. 
Starting from a separation of 0.1~pc, a combination of disk-driven 
migration and inspiral due to gravitational radiation leads to 
merger within $2 \times 10^7 \ {\rm yr}$. 

In Figure \ref{fig3} we show the disk surface 
density profile during the migration and inspiral. There are 
two phases. In the first, the secondary migrates inwards within 
a gap under the action of disk torques, while the inner disk is 
partially depleted by accretion onto the primary black hole. The 
accretion rate onto the primary black hole in this first phase is 
somewhat reduced compared to the value it would have in the absence 
of a binary. This behavior can be easily understood -- adding a 
binary companion to the disk merely increases the amount of angular 
momentum that must be transported outwards before mass can be 
accreted. Flux-limited samples of AGN may therefore  
select {\em against} finding close binaries. Disk-driven migration 
continues until angular momentum loss due to gravitational radiation 
first becomes important, at a separation of $\sim 10^{-2} \ {\rm pc}$. 
Thereafter, the inspiralling black hole begins to sweep up the gas 
in the inner disk, forming a narrow inner spike in the disk surface 
density that is pushed inwards. The outer disk is unable to evolve 
on such a rapid timescale, so a large gap develops between the outer 
disk and the secondary.

The final stages of the merger drive rapid accretion of the inner disk. 
Around $3 \times 10^5 \ M_\odot$ of gas remain in the inner disk 
when the secondary reaches $10^{-3} \ {\rm pc}$, around 200~yr 
prior to merger. This implies an enormous accretion rate (formally 
exceeding $10^6 \ M_\odot {\rm yr}^{-1}$) immediately prior to 
merger, and signals the breakdown of the thin-disk approximation. 
Instead of being radiated, the dissipated energy will instead 
go into thermal energy, puffing up the inner disk and forming a 
hot accretion flow \citep{begelman82}. There are large 
uncertainties in the further evolution, but we speculate that 
the final merger is likely to occur within a common envelope 
formed from the disrupted inner disk. As in the common 
envelope evolution of binary stars \citep{taam78}, strong outflows are 
extremely likely, with characteristic velocities comparable 
to the orbital velocity of $\sim 10^4 \ {\rm kms}^{-1}$.
Strong outflows from galactic nuclei may therefore act as 
signposts to systems where binaries are either currently 
inspiralling, or have recently merged.

\subsection{Quiescent galactic nuclei from black hole binaries}

The timescale on which black holes form a hard binary, 
following a galactic merger, depends upon the stellar 
dynamics of the interaction between the binary and the 
stellar population. Gas could well flow into the merged nucleus 
on a shorter timescale \citep{hernquist95}.
As a result, even if the merging galaxies contain substantial 
quantities of gas, it is not obvious that the resulting black 
hole binary will find itself in a gas-rich environment. 

\begin{figure}
\plotone{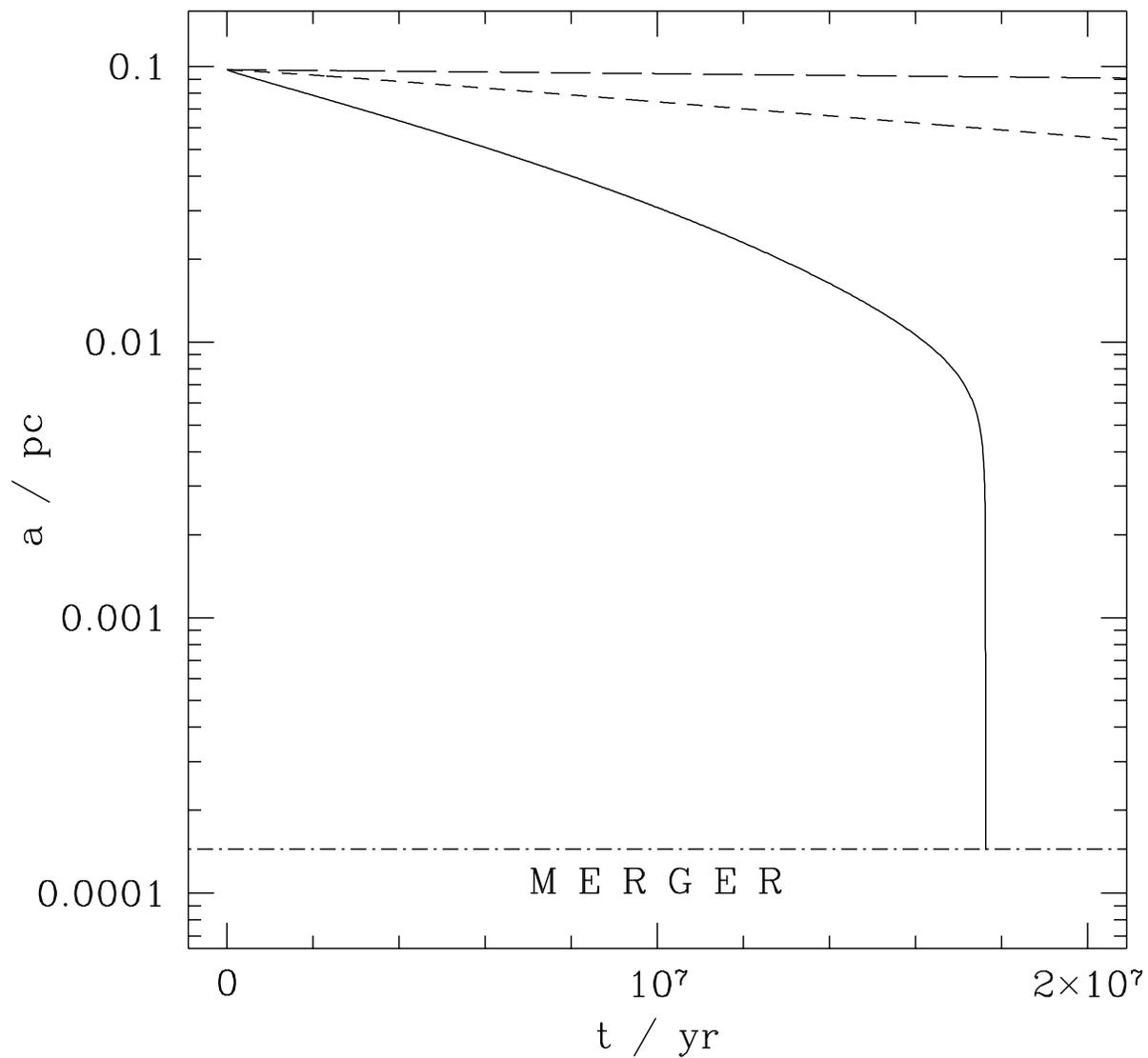}
\caption{Migration of the secondary black hole in disks with initial accretion 
	rates of $1 \ M_\odot {\rm yr}^{-1}$ (solid line), $0.1 \ 
	M_\odot {\rm yr}^{-1}$ (short-dashed line) and $10^{-2} \ 
	M_\odot {\rm yr}^{-1}$ (long-dashed line). Binaries embedded within 
	relatively massive disks are quickly driven into the gravitational wave 
	dominated regime and merge promptly.} 
\label{fig2}
\end{figure}

\begin{figure}
\plotone{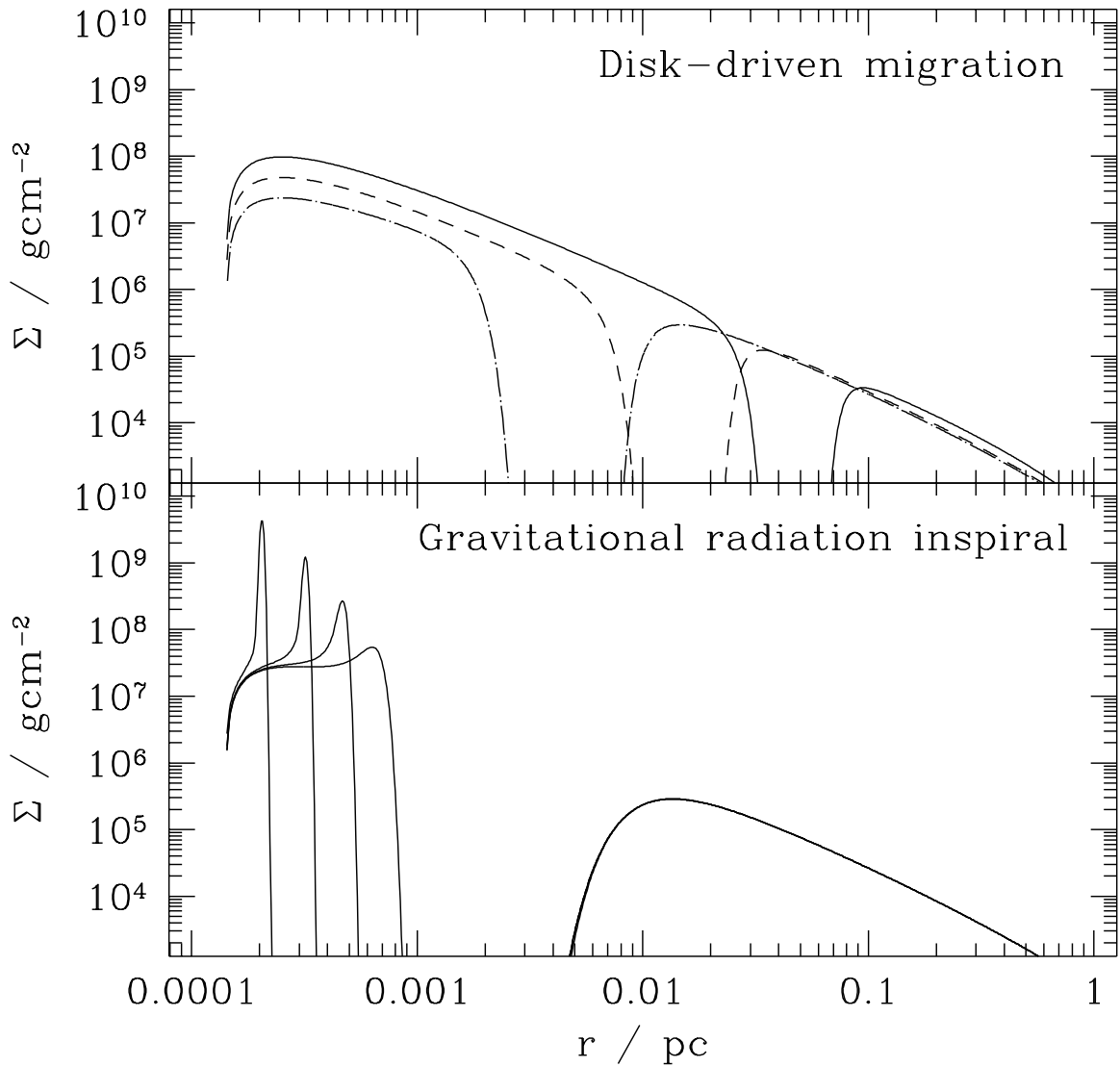}
\caption{Disk surface density during black hole migration. The black hole 
	migrates inwards within a gap during the disk dominated phase (upper 
	panel, curves show $\Sigma$ at three epochs). Once orbital decay due 
	to gravitational radiation dominates, 
	the inner disk is swept up by the inspiralling hole and forms a surface 
	density spike (lower panel). The outer disk, which is unable to evolve on such 
	a short timescale, maintains a fixed surface density profile.} 
\label{fig3}
\end{figure}

Figure \ref{fig2} also shows how the secondary migrates in disks that 
have the same viscosity as that considered previously, but initially 
lower disk masses and accretion rates. We ran one model with an initial 
accretion rate of $0.1 \ M_\odot {\rm yr}^{-1}$ and a 
disk mass of $6 \times 10^6 \ M_\odot$, and a second with 
$\dot{M} = 10^{-2} \ M_\odot {\rm yr}^{-1}$ and 
$M_{\rm disk} = 6 \times 10^5 \ M_\odot$. Migration rates in these 
lower mass disks are substantially reduced, so it is possible that 
there could be a long-lived population of binaries embedded in low mass 
disks. The reduced level of accretion onto the binary, 
due to the gravitational torques, could then contribute to the 
surprisingly low levels of accretion inferred in most local galactic 
nuclei. This suppression of accretion, however, is only 
effective for cold gas. Hot gas, with low specific angular 
momentum, can surmount the angular momentum barrier and replenish 
small disks around the binary components \citep{bate1997}. 
Explaining why this accretion does not lead to large luminosities 
\citep{fabian95,loewenstein01} almost certainly involves the radiation 
physics \citep{narayan95}, and possibly 
hydrodynamics \citep{blandford99}, of low $\dot{M}$ flows. 

\section{Summary}

The ubiquity of supermassive black holes in the local Universe \citep{magorrian98} 
suggests that black holes may also be common in the high redshift progenitors 
that merged to make up present day galaxies. Although a large black hole 
fraction at high-$z$ is not proven \citep{menou2001}, a high frequency of 
mergers between galaxies harboring black holes would  
have interesting implications for the 
accretion history of AGN \citep{haehnelt1998,haehnelt2000,cv2000}, and for 
the structure of galaxies themselves \citep{ravindranath2001}. 

In this {\em letter}, we examined the coupled evolution of a supermassive 
black hole binary embedded in a gaseous accretion disk. First, we showed  
that, in a gas-rich environment, a binary with a separation of $0.1 \ {\rm pc}$ 
merges within around $10^7 \ {\rm yr}$. Close binary quasars are therefore 
expected to be uncommon. Long lived black 
hole binaries are more 
likely to be found in apparently quiescent galactic nuclei at low redshift, 
where the disk accretion rate is lower, and where migration may be further 
slowed by inefficient disk angular momentum transport \citep{menou01}.
Second, we examined the possible observational signatures of the 
merger. We argued that large enhancements to the accretion rate were 
unlikely during the phase of disk-driven migration. Forced 
accretion of any inner disk during the final, gravitational wave driven inspiral, 
however, is likely to lead to much of the inner disk 
being expelled in a high velocity quasi-spherical outflow. 

\acknowledgements

PJA thanks Yale for hospitality during the 
course of this work. Numerical simulations made use of the 
UK Astrophysical Fluids Facility.

\clearpage

\end{document}